\documentclass[amssymb,aps,showpacs,prl,amsmath]{revtex4}

\usepackage{graphicx}
                   
\usepackage{latexsym}

\newcommand{\be}{\begin{equation}}
\newcommand{\ee}{\end{equation}}

\newcommand{\bex}{\begin{eqnarray}}
\newcommand{\eex}{\end{eqnarray}}
\newcommand{\bmin}{\begin{center}\begin{minipage}{460pt}}
\newcommand{\emin}{\end{minipage}\end{center}}
\begin{document}

\title{Computable Functions, the Church-Turing Thesis
 and the Quantum Measurement Problem}
\author{R. Srikanth}
\email{srik@rri.res.in}
\affiliation{Optics Group, Raman Research Institute
Bangalore- 560~080, Karnataka, India.}


\begin{abstract}
It is possible in principle to construct quantum mechanical observables and 
unitary operators which, if implemented in physical systems as measurements
and dynamical evolution, would contradict the Church-Turing thesis, which lies
at the foundation of computer science. Elsewhere we have argued that the quantum
measurement problem implies a finite, computational model of the 
measurement and evolution of quantum states. If correct, this approach helps to identify the
key feature that can reconcile quantum mechanics with the Church-Turing thesis:
finitude of the degree of fine-graining of Hilbert space. This suggests that the Church-Turing 
thesis constrains the physical universe and thereby it highlights a surprising connection between 
purely logical and algorithmic considerations on the one hand and physical reality on the other.
% We next ask the question: what is
% the origin of this phenomenology? An infinitely
% fine Hilbert space, being complete in a G\"odel sense, would be equivalent to a
% too powerful model of computation to be consistent. The requirement of consistency
% leads to finitude of resolution of Hilbert space as a guarantee of incompleteness.
% According to this line of reasoning, the classicality of macro-systems is a
% manifestation of the G\"odel incompleteness theorem.
\end{abstract}
\maketitle

Quantum mechanical measurements on a physical system are represented by 
observables- Hermitian operators on the state space of the observed system.
It is an important question whether all observables and all unitary
evolutions may be realized, in
principle on physical systems \cite{niels}. 
It turns out that ideas from computer
science are crucial to the analysis. The reason is that some problems
that are classically considered uncomputable can sometimes be re-cast
as appropriate quantum observables or unitary operations, which, if
phyiscally implementable, could apparently render them computable.
There has appeared to be no known principle that forbids
the implementation of such quantum mechanical observables or 
unitary dynamics. 
If feasible, they would imply that quantum mechanics not only impacts computational
complexity, in terms of speeding up certain computations \cite{shor}, 
but also impacts {\em computability}. 
In a remarkable work that founded modern computer science, Turing \cite{tur,stm} defined
a class of functions now known as recursive or computable functions. The 
{\em Church-Turing thesis} \cite{cop}
of computer science states that these functions correspond
precisely to any {\em effective} or {\em mechanical} method for obtaining the values
of a mathematical function. The Church-Turing thesis is fundamental to theoretical
computer science, since it asserts that the mathematical class of functions studied
by computer scientists, the computable functions, is the most general class of
functions that may be effectively calculated by any physical means. It is an empirical
statement, not a mathematical theorem, that has been verified over 65 years of
testing \cite{hofs,stm}.

The present work reexamines
the question of whether, accepting the Church-Turing thesis as an informal
guiding principle for physics \cite{svo97}, we can identify a principle that can reconcile this
conflict between it and the general model
of computation that quantum mechanics seems to make possible.
Naturally, for specific systems particular measurements and unitary dynamics
could be precluded by some system specific features, such as superselection
rules. We will not be concerned with such specific features, but rather with
general considerations. In fact, we find that
we cannot be confined to the physical realm of measurement and dynamics, 
but are led to examine the algorithmic foundations of physical processes,
in the light of the computational model of quantum measurement \cite{srik}.

For convenience, we refer to computable partial functions of a single non-negative
integer as {\em programs}. The qualification `partial' signifies that there may
be some input values for which the function is not defined. An example of a program
is the function $f(x) = x^3$, which may be computed using a suitable mechanical
procedure (a Turing machine). On the other hand, an example of a function that
does not stop or {\em halt} for some input, and is thus undefined for that input,
is $f(x) \equiv$ ``the smallest number that is the sum of $x$ squares", which will
not halt on $x=4$ \cite{legr}. 

The question of whether or not a given program, numbered $p$ (according to any
consistent enumeration scheme, such as G\"odel numbering) halts on input $i$ is obviously
a question of considerable practical importance: we would like to know
whether a given algorithm will terminate or not. To understand this better,
Turing defined the {\em halting function} $h(p,i)$ as some algorithm to decide
whether program $p$ halts on input $i$ and demonstrated using
a {\em reductio ad absurdum} argument that the halting problem is not a decidable.
A version of this argument is as follows. Suppose we define the halting function by:
\be
\label{eq:turing}
h(p,i) = \left\{\begin{array}{ll}
	= 0~ & {\rm if~}p(i){\rm ~loops~endlessly,} \\ 
        = 1~ & {\rm otherwise},
 \end{array}\right. 
\ee
and diagonal function $r(n)$ by:
\be
r(n) = \left\{\begin{array}{ll}
         0~ & {\rm if~} h(n,n) = 0,  \\ 
         {\rm loops~endlessly} & {\rm ~otherwise}.
         \end{array}\right.
\ee
Suppose $r(n)$ is computable and this can be ascertained by running $h(r,n)$. Yet,
setting $n=r$ above, we find $r(r) = 0 \Leftrightarrow h(r,r) = 0
\Leftrightarrow r(r)$ diverges, which is a clear contradiction. Thus, $r(n)$ cannot
be computable. Since the computability of $h(p,i)$ would imply that of $r(n)$,
it follows that the halting function is also noncomputable here.
This is related to G\"odel's incompleteness theorem that says that if a
consistent axiomatic system is rich enough to be self-referential, then 
it will be incomplete in that true but unprovable statements can be made in
it \cite{God}.

It is reasonable to expect that quantum computation does not impact computability,
since the latter appears to follow mostly from logical considerations rather than
any model of computation \cite{bern97}. Moreover, one might expect that
noncomputability might be embedded right in the (classical) specification 
of the implementation of such otherwise hypercomputational operations. 
A quantum algorithm can in principle side-step this 
contradiction, by distinguishing quantum and classical algorithms.
A {\em quantum} function $qh(p,i)$, similar to Eq. (\ref{eq:turing}), can conceivably
exist to determine whether the classical program will halt on input $i$ or not. The
above contradiction would be avoided if $qh(p,i)$ cannot
take as argument the modified program $r(n)$, which is now of {\em quantum}
character because it calls $qh(p,i)$ as subroutine \cite{kieu03}.
This follows from the fact that $qh(p,i)$ can accept only integers
(enumerations of programs, such as G\"odel numbering), 
while quantum programs have to be labelled
by real numbers, which have a larger cardinality than integers. 
The idea is that the computer should not be powerful enough to compute its
own diagonal function, which if it is, leads to the above contradiction.
However, this suggests that a quantum algorithm can in principle compute the 
halting problem for Turing machines.

Let us consider Hilbert's tenth problem, the Entscheidungsproblem (decision
problem), which asks for a procedure to decide whether a given Diophantine equation
(polynomial with integer coefficients) $D(x)$ has a solution in the domain 
of natural numbers, i.e., $x \equiv (x_1,\cdots,x_n) \in \mathbb{N}^n$. 
This is known to be Turing uncomputable \cite{h10}. 
However, suppose we define a `Diophantine evolution', $U_D$, by:
\be
\label{eq:diophe}
U_D \equiv \sum_{x=0}^{\infty}|D^2(x),x\rangle\langle x|,
\ee
where $\{|x\rangle\}$ is an orthonormal basis for the state space of some
physical system with a countably infinite dimensional state space.
Here $|D^2(x),x\rangle$ is a state with energy eigenvalue $D^2(x) 
\equiv (D(x))^2$. It is easy
to verify that $U_D$ is unitary. We note that $U_D$ is equivalent to measuring a 
`Diophantine observable' $\hat{\cal D}$:
\be 
\label{eq:diopho}
\hat{\cal D} \equiv \sum_{x=0}^{\infty}D^2(x)|x\rangle\langle x|,
\ee
If one can design a Hamiltonian that effects $U_D$, then the {\em search}
for a solution to $D(x)$ is equivalent to {\em checking} whether the ground state 
of the Hamiltonian has energy $E_g=0$. An exhaustive search can never terminate
if no solution exists, which is a manifestation of the undecidability of
the Entscheidungsproblem. However,
in the quantum case, if the required Hamiltonian can be prepared, and its
ground state accessed in finite time, then a {\em single} check suffices to
decide whether $D(x)$ has a solution \cite{kieu03}. 
It is clear that the success of the above method depends on superposition
in Fock space and the guarantee that the ground state minimizes energy.
The power of the ground state
oracle derives from ``energy tagging", which appears to be more powerful than
a usual quantum search oracle.

If we may abstract the computational model that this quantum procedure 
embodies, devoid of any
physical content, and idealize it, the result is what may be called an
{\em infinite Turing field} ${\cal T}(\aleph_0,\aleph_0)$: 
the set of countably infinite ($\aleph_0$) Turing machines
$T_x$ simultaneously evaluating $D(x)$, a total recursive function,
on each possible value of $x$, having
infinite memory, and the power to communicate instantaneously 
with each other (i.e, message passing rate = $\aleph_0$/sec),
such that the moment some $T_x$
finds $D(x)=0$ on its $x$, all $T_x$'s are brought
to halt, and, deterministically, precisely this value $x$ and no other 
is printed out. The twin arguments of ${\cal T}(\aleph_0,\aleph_0)$
signify that it is composed of infinitely many Turing machines that are able
to talk to each other infinitely fast. In modern computer terminology, this is
akin to a computer having infinite RAM and infinite clockrate.
Note that this is a well-defined model because the set of all
possible values of $n$-tuples $x \in \mathbb{N}^n$ has the same cardinality ($\aleph_0$)
as the set of natural numbers $\mathbb{N}$. An alternate definition of 
${\cal T}(\aleph_0,\aleph_0)$
is that it is a Turing machine that performs calculations infinitely fast (i.e,
at the rate of $\aleph_0$ logical
operations per second, or ops). ${\cal T}(\aleph_0,\aleph_0)$ 
transcends the Turing barrier by performing infinite calculations in finite time.
Further, by stipulating that it accepts only problems of finite size, we ensure
that it is not powerful enough to compute its own diagonal function and thus
remains logically consistent for our present needs.  % \cite{God} 
The proposed quantum algorithms for computing
Turing uncomputable functions \cite{kieu03,Krist} are in effect 
emulating a probablistic version of ${\cal T}(\aleph_0,\aleph_0)$, where the
level of confidence can be made arbitrarily high as time increases. 
They involve a countably infinite dimensional Hilbert space,
where each dimension in the generic quantum superposition
corresponds to a Turing machine in the field ${\cal T}(\aleph_0,\aleph_0)$. 

A demonstration that a quantum computer, even probablistically, emulates 
${\cal T}(\aleph_0,\aleph_0)$ would
violate the Church-Turing thesis, for it would demonstrate an effective
algorithm to compute a function that is Turing uncomputable. There is an
implicit and quite reasonable assumption on which the Turing barrier transcendence 
attributed to quantum algorithms rests: the presumed
infinitude of the universe, in particular, of Hilbert space.  
However, the computational model for quantum measurement 
(CMQM) \cite{srik} suggests that, insofar as 
the quantum state dynamics of physical systems may be regarded as 
information registered and computations performed
by the (quantum) universe, the latter is
equivalent to a {\em finite} Turing field: a finite number of Turing machines
interlinked by finite speed message passing. If true, then this furnishes a very basic
algorthmic-physical principle why quantum computers cannot break
through the Turing barrier. What is the evidence for this principle?
In Ref. \cite{srik}, we argue that {\em the quantum measurement
problem} is the required evidence.

A simple recapitulation of the measurement problem is as follows. Suppose we have
a spinor prepared in the state $|+\rangle_x = (1/\sqrt{2})(|+\rangle_z
+ |-\rangle_z)$ and a macroscopic meter that measures the observable
$\hat{\sigma}_z$. The meter has three macroscopic states, given by $\{|+\rangle_M,
|-\rangle_M, |0\rangle_M\}$. If the spinor exists in one of the $\hat{\sigma}_z$
eigenstates, the meter gives a measurement according to the interaction:
$|\pm\rangle_z|0\rangle_M \longrightarrow |\pm\rangle_z|\pm\rangle_M$. Yet,
when the state $|+\rangle_x$ is measured, this does not lead to a superposition
according to $|+\rangle_x|0\rangle_M \longrightarrow
(1/\sqrt{2})(|+\rangle_z|+\rangle_M + |-\rangle_z|-\rangle_M)$, as the principle
of linearity suggests, but probablisitically to {\em one} of the outcomes
$|\pm\rangle_z|\pm\rangle_M$. The absence of macroscopic superpositions, that one
would naturally expect during measurement via system-apparatus
entanglement, is the measurement problem.

In CMQM, the dynamical evolution of physical systems are
modeled as computational acts on an abstract, finite, algorithmic structure underlying
physical reality. The apparent break-down in linearity noted above and the 
mesoscopic threshold at which it presumably occurs are related to a finite degree
of fine-graining of the Hilbert space of physical systems (in general, of the universe).
The fine-graining parameter $\mu$ (in bits per amplitude; $\mu/2$ bits each
specifying the real and the imaginary part) specifies the precision
to which states, and their unitary evolution, can be specified in principle. 
According the `principle of computational resolution' enunciated in the model, 
$\mu$ should be large enough to support a uniform superposition. 
That is, in a system of $n$ qubits,
each amplitude in a uniform superposition $\sum_{j=0}^{2^n-1}\lambda_j|j\rangle$
is given by $|\lambda_j| = 2^{-n/2}$, which should be greater than $2^{-\mu/2}$,
so that (roughly) $n \le \mu$. 
If a system, through entanglement engendered by interaction with its environment, becomes
larger than is this, i.e., the dimension of its Hilbert space exceeds $2^{\mu}$,
the system is said to become computationaly unstable because
the computational and mnemonic support for its physical evolution is overwhelmed, and
to discontinuously suffers an information transition
to a separable state chosen probablistically. 
Such transitions in the abstract memory registers are identified
with the `state vector reduction' or `collapse of the wavefunction'.  Subsequent
interaction again drives the exponential enlargement of the system's state space
via entanglement with its environment. 
The continual cycle of alternating dynamical evolutions
and information transitions is responsible for the appearance of classicality at the
macro-level, which is described by the Lindblad equation \cite{lind}. 
Therefore, $\mu$ determines the Heisenberg
cut, the mesoscopic threshold that presumably separates the quantum microcosm from
the classical macrocosm. If no such cut existed, CMQM implies that macroscopic
superpositions would be ubiquitous, which they aren't
(for comments on the important role of decoherence, cf. \cite{srik}).
In short, in this light we infer that the classicality of the macroscopic world 
implies a finite computational and memory
support of the universe for the evolution of physical systems, equivalent to a 
{\em finite Turing field} ${\cal T}(M,V)$, where $M, V < \aleph_0$.

In the case where the large dimension arises from accessible excited states, rather than
from interactions, the finite value of $\mu$ implies an upper bound
to number of basis states that can be evolved coherently in a superposition,
given by about $2^{\mu}$, because amplitudes that are in principle magnitudinally
much smaller than $2^{-\mu/2}$ will be re-set to zero, with concommitant loss of
information from the state. If the number of `important' superposed states is
sufficiently small, which is the situation in most cases, this truncation effect
is hardly signifcant. However, when the evolution passes through intermediary 
states where amplitude can be spread out among a large number of states,
approaching the above bound, the deviation from unitary evolution can be significant. 
The apparent behaviour of the system will be one whose unitary evolution is
interrupted because of decoherence or of having been
measured. In a complicated energy landscape of states, the system can collapse into
a local minimum. The evolution would no longer be given by an equation of the
form Eq. (\ref{eq:diophe}), but of a more general positive operator that leads
to a final state that is mixed.

We can use observations to bound $M$ and $V$ from above.
If $S$ represents the entropy
of an isolated system, then CMQM implies (roughly) that the memory
corresponding to it (ie., the total information required to specify its 
microstate in ${\cal H}_{\mu}$, the finitely fine-grained Hilbert space)
is about $M \equiv e^{S/k_B}\mu$ and total speed of computation
that corresponds to sustaining the evolution of the system is about
$V = 2^{\mu/2}e^{S/k_B}E/\hbar$ ops, where $E$ is energy of the
system. Suppose for instance that the largest object whose unitary evolution 
(run at $\mu$ bit precision) can be stably supported by the universe is 
an Avogradro number of qubits, beyond which the system becomes computationally
unstable and hence classical. Then $\mu$ can be estimated to be about
$10^{23}$ bits, from which one can estimate $M 
\approx 2^{10^{23}}10^{23} \approx 2^{10^{23}}$ bits and 
$V \approx 2^{1.5 \times 10^{23}}E/\hbar \approx 2^{1.5\times 10^{23}}$ ops 
(for more realistic estimates, cf. Ref. \cite{srik}).
Not surprisingly, the memory overhead and the computational speed of all conventional 
computers combined is far lesser. With 
about $10^9$ computers equipped with terabit memory and
operating at a clock rate of about $10^9$ Hz
performing about $10^5$ elementary logical operations per clock cycle, all the
human-made computers in the world can store $10^{21} \ll M$ bits, and
are operating at no more than $10^{23} \ll V$ ops.

If the evolution of the universe is equivalent to a finite algorithmic procedure, 
then it is clearly
subject to the limitations of a Turing machine, even though it is much more powerful
and faster than any conventional computer that can be created. In particular,
the finitude of $\mu$ means that all possible quantum programs
that correspond to physical reality, including $qh(p,i)$ and $r(n)$, 
can be enumerated at that digital precision. 
The contradiction that arose in Eq. (\ref{eq:turing}) can no longer
be avoided, and we recover the Church-Turing thesis. Similarly, according to the
computational resolution principle, a superposition of more than 
$2^{\mu}$ states cannot be coherently maintained without interruptions to its
unitary evolution.
In practice this means that Diophantine equations
with possible solutions larger than $2^{10^{23}}$ are effectively uncomputable using $U_D$ in
Eq. (\ref{eq:diophe}). Suppose all
the degrees of freedom in the universe were available for storing information:
this can be estimated to be about $10^{120}$ bits, including gravitational degrees of
freedom \cite{llo02}. Even if we grant $\mu=10^{120}$ so that coherent evolution
of the universe as a whole can be computationally supported,
CMQM implies that the number of $T_x$'s in a finite Turing field equivalent to
the quantum algorithm cannot exceed $2^{10^{120}}$; in an arbitrary basis,
the number of elements in $U_D$ in Eq. (\ref{eq:diophe}) cannot exceed 
$2^{10^{120}} \times 2^{10^{120}}$. This seems quite unlikely even as the system is
`rotated' over vast tracts of states, corresponding to various potential solutions
to $D(x)$. Accordingly, the possible largest solutions to a non-recursive
function (such as the solution to an arbitrary Diophantine equation) that can  
be explored are bounded above by $2^{10^{120}}$. One might say that
there is not enough room in the resolution limited Hilbert space 
${\cal H}_{\mu}$ of any quantum computer
to explore all of the solution space, which is $\mathbb{N}^n$.

The point can be further clarified in taking a spin $\frac{1}{2}$ system.
Consider 
\be \Omega \equiv \sum_{p {\rm ~halts}}2^{-|p|},
\ee
Chaitin's constant, which is
the probability that an arbitrary program will halt, where $|p|$ is
the size (in bits) of program $p$ \cite{chaitin}. Knowing the binary expansion of $\Omega$
is equivalent to solving the halting problem for all $p$.
If the universe, and thence the Hilbert space that describes it, are infinite,
one might consider an effective method of designing the evolution $U_C \equiv 
\exp(-i\Omega\sigma_x)$. Applying the dynamics $U_C$ on a system 
prepared in the $|\uparrow\rangle_z$ state (spin up in the $z$ direction)
results in the state:
\be
\cos(\Omega)|\uparrow\rangle_z + \sin(\Omega)|\downarrow\rangle_z.
\ee
By repeatedly performing this procedure and measuring $\sigma_z$, we may
determine $\cos\Omega$ and thus $\Omega$ to any desired accuracy, with
arbitrarily high confidence. According to the computational model,
however, $\cos\Omega$ can never be known with precision greater than
$\mu$. Although $\mu$ can be arbitrarily large, and
thus in principle $\Omega$ calculated in principle to arbitrary precision,
in any given instance of the universe, $\mu$ is finite and thus $\Omega$
can never truely be computed. 

This work has discussed whether the observables and dynamics permitted by
quantum mechanics permit quantum computation to
transcend the Turing barrier that bounds computable functions.
If they do, they would in principle contradict the Church-Turing
thesis. Using concepts from computer science and in the framework of CMQM,
we resolve this conflict 
by pointing out that the measurement problem
implies that a quantum computer corresponds to a finite Turing field. % , the array of a
% finite number of Turing machines endowed with finite communication speed.
This viewpoint, which is based on a realistic interpretation of the state vector
as a not entirely accessible object that resides and evolves in an abstract space, 
is referred to as `informational realism'. It is perhaps the simplest
and most direct interpretation of the basic mathematical formalism. Its physical
content shows up only when the large but finite upper bound it imposes on the dimension 
of coherently evolving systems is reached.


\begin{thebibliography}{100}
% \bibitem{dirac} P. A. M. Dirac, {\em The Principles of Quantum Mechanics}
% (Oxford University Press, Oxford, 1958), 4th ed. p37.
\bibitem{niels} M. A. Nielsen, Phys. Rev. Lett. {\bf 79} 2195, (1997).
\bibitem{shor} P. Shor, SIAM Journal of Computing {\bf 26}, 1484 (1997);
L. K. Grover, Phys. Rev. Lett. 79, 325 (1997).
\bibitem{tur} A. Turing, Proc. Lond. Math. Soc. (ser 2.) {\bf 42}, 230 (1937); 
(a correction) {\bf 45}, 161 (1937).
\bibitem{stm} R. Penrose, {\em The Emperor's New Mind} (Oxford University Press, 1989);
{\em Shadows of the Mind} (Vintage, 1995). 
\bibitem{cop} B. J. Copeland, {\em The Church-Turing Thesis}, 
The Stanford Encyclopedia of Philosophy (Fall 2002 Edition), Edward N. Zalta (ed.), URL =
{\tt <http://plato.stanford.edu/archives/fall2002/entries/church-turing/>}
\bibitem{hofs} D. R. Hofstadter, {\em G\"odel, Escher, Bach: An Eternal Golden
Braid} (Basic Books, New York, 1979).
\bibitem{svo97} K. Svozil, in {\em Unconventional Models of Computation},
eds. Cristian S. Calude, J. Casti and MJ Dineen ( Springer-Verlag, Singapore, 1998); 
eprint quant-ph/9710052.
\bibitem{srik} R. Srikanth, Quantum Information Processing {\bf 2}, 153 (2003).
\bibitem{legr} According to a theorem due to Legendre, all natural numbers can be 
expressed as the sum of four squares. Eg., $7 = 2^2 + 3\times1^2$.
\bibitem{God} K. G\"odel, Monatschrifte f\"ur Mathematik und Physik {\bf 38}, 173 (1931). 
\bibitem{bern97} E. Bernstein and U. Vazirani, {\em Quantum Complexity Theory}
SIAM J. Comput. {\bf 26}, 1411 (1997).
% \bibitem{svo98} K. Svozil, in {\em Unconventional Models of Computation},
% eds. C. S. Calude, J. Casti and M. J. Dineen (Springer Verlag, Singapore) pp. 371-385.
% \bibitem{cot03} P. Cotogno, Brit. J. Phil. Sci. {\bf 54}, 181 (2003).
\bibitem{kieu03} T. D. Kieu, Intl. J. Theor. Phys. {\bf 42} 1461 (2003);
Contemporary Physics {\bf 44}, 51 (2003).
\bibitem{h10} Y. V. Matiyasevich, {\em Hilbert's Tenth Problem} (MIT Press, 1993).
\bibitem{Krist} C. S. Calude and B. Pavlov, 
Quantum Information Processing {\bf 1}, 107 (2001); eprint quant-ph/0112087.
\bibitem{lind} G. Lindblad, Commun. Math. Phys. {\bf 48}, 119 (1976).
\bibitem{llo02} S. Lloyd, Phys. Rev. Lett. 88 (2002) 237901.
\bibitem{chaitin} G. Chaitin, {\em Algorithmic Information Theory} (Cambridge 1987);
E. W. Weisstein, MathWorld (Wolfram Web Resource):
{\tt http://mathworld.wolfram.com/ChaitinsConstant.html}
\end{thebibliography}
\end{document}